\documentclass[fleqn,usenatbib]{mnras}

\usepackage{newtxtext,newtxmath}

\usepackage[T1]{fontenc}

\DeclareRobustCommand{\VAN}[3]{#2}
\let\VANthebibliography\thebibliography
\def\thebibliography{\DeclareRobustCommand{\VAN}[3]{##3}\VANthebibliography}

\usepackage{graphicx}	
\usepackage{amsmath}	
\usepackage{multirow}

\usepackage{graphicx}
\usepackage{ulem}
\usepackage{xcolor}

\title[SGRB-jet through magnetized media]{Dynamics of a relativistic jet through magnetized media}

\author[Garc\'ia-Garc\'ia et al.]{Leonardo Garc\'ia-Garc\'ia$^{1}$\thanks{E-mail: lgarcia@astro.unam.mx}, Diego L\'opez-C\'amara$^{2}$, Davide Lazzati$^{3}$
\\
$^{1}$Instituto de Astronomía, Universidad Nacional Autónoma de México\\
$^{2}$C\'atedras CONACyT -- Universidad Nacional Aut\'onoma de M\'exico, Instituto de Astronom\'ia, AP 70-264, CDMX  04510, M\'exico\\
$^{3}$Department of Physics, Oregon State University, 301 Weniger Hall, Corvallis, OR 97331, U.S.A.}

\date{Accepted XXX. Received YYY; in original form ZZZ}

\pubyear{2021}

\begin{document}
\label{firstpage}
\pagerange{\pageref{firstpage}--\pageref{lastpage}}
\maketitle

\begin{abstract}
The merger of two neutron stars (NSs) produces the emission of gravitational waves, the formation of a compact object surrounded by a dense and magnetized environment, and the launching of a collimated and relativistic jet, which will eventually produce a short gamma-ray burst (SGRB). The interaction of the jet with the environment has been shown to play a major role in shaping the structure of the outflow that eventually powers the gamma-ray emission. In this paper, we present a set of 2.5 dimensional RMHD simulations that follow the evolution of a relativistic non-magnetized jet through a medium with different magnetization levels, as produced after the merger of two NSs. We find that the predominant consequence of a magnetized ambient medium is that of suppressing instabilities within the jet, and preventing the formation of a series of collimation shocks. One implication of this is that internal shocks lose efficiency, causing bursts with low-luminosity prompt emission. On the other hand, the jet-head velocity and the induced magnetization within the jet are fairly independent from the magnetization of the ambient medium. Future numerical studies with a larger domain are necessary to obtain light curves and spectra in order to better understand the role of magnetized media.
\end{abstract}


\section{Introduction}
\label{s:intro}
Gamma-ray Bursts (GRBs) are high-energy transitory events of $\gamma$-rays \citep{klebesadel73, meszaros2019}. GRBs have an isotropic distribution\footnote{See for example the distribution of the GRBs detected by CGRO-Batse https://heasarc.gsfc.nasa.gov/docs/cgro/batse/}, are produced at cosmological distances \citep{costa97}, and are classified mainly into two populations according to their duration \citep{1993K}. Events that last $T_{\rm 90}\lesssim 2$~s are classified as short GRBs (SGRBs), and those with duration $T_{\rm 90}\gtrsim2$~s as long GRBs. The SGRB population is characterized by isotropic energies of E$_{\rm iso}\sim10^{50}-10^{52}$~erg \citep{2009Gh}, isotropic luminosities of L$_{\rm iso}\sim10^{50}-10^{53}$~erg~s$^{-1}$ \citep{ghirlanda2009}, velocities of order $\Gamma\sim$20-10$^{3}$ \citep{Ghi2018}, and redshifts of $<z>\sim$0.5 \citep{2014Ber}.

SGRBs are produced by powerful relativistic and collimated jets result of the coalescence of a binary neutron star (NS) system e.g \citep{AbbotA2017, AbbotB2017,Goldstein2017,Hallinan2017,Margutti2017,Troja2017,Lazzati2017,Lazzati2018,Mooley2018,Ghirlanda2019}. The SGRB-jets have opening angles of $\theta_{j}\sim5^{\circ}-25^{\circ}$ \citep{2015Fong, rouco2022}, and thus, have jet luminosities of L$_{\rm j} \sim 10^{47}-10^{51}$~erg~s$^{-1}$.

During the merger of the NSs dense material is launched (termed as ``dynamical ejecta''), and a newly formed compact object (either a high mass NS or a black hole, BH) is surrounded by an accretion disk \citep{om2022}. Two bipolar jets, which are launched from the central engine and evolve through the dynamical ejecta, have internal shocks that convert a fraction of the total energy to radiation \citep{Daigne1998,Bo2009} and need to reach distances larger than the photospheric radius ($R_{\rm ph} \sim 10^{12}-10^{13}$~cm) to be optically thin and emit the highly-energetic photons \citep[photospheric model,][]{fireball,Rees2005,Peer2006,Giannios2007,Lazzati2009,Ryde2011}.
While the jet is optically thick ($d<R_{\rm ph}$), the dense ejecta may produce the formation of different components in the jet (or surrounding it). First, the a shock is produced at the front of the jet \citep[jet-head, jh,][]{Blandford1974,Scheuer1974}. Second, the shocked material that the jh pushes sideways may produce a hot bubble \citep[cocoon,][]{Begelman1989}. Third, the large pressure of the cocoon may collimate the jet and form recollimation shocks (RSs) within the jet \citep{om2022}.

Previous studies have shown that the lighcurve (LC) is highly dependent on the jet structure \citep{Duffell2018,Lazzati2018,Mooley2018,Nakar2018,Ger2021}, thus, it is necessary to understand how the jet is shaped close to the central engine in order to interpret the obtained LCs. The medium expelled during the merger of the NSs has a crucial role in shaping the jets (which are launched form the newly formed NS or BH). The importance of the medium has been studied by following the propagation of a relativistic outflow through a medium with different density, pressure, velocity profiles and total mass in the domain through relativistic hidrodynamical simulations in two or three dimensions \citep{Aloy2005, Bucciantini2008, Bucciantini2009, Davide2013, Duffell2015, Murguia2014, Nagakura2014, Murguia2017, Bromber2018, Duffell2018, Granot2018, Harrison2018, Lamb2018, Lazzati2018, Xie2018, Got2018a, Got2018b, Gill2019,LazzatiPerna, Lazzati2020,Hamidani2020, Ger2021, hamidani2021, Murguia2021}. Moreover, \citet{Got2020,Got2021a,Got2021b} simulated highly, intermediate and low-magnetized relativistic jets evolving through a dense medium using relativistic magneto-HD (RMHD) simulations and found that magnetized jets are more stable. The medium through which the relativistic jets drills, though, have not been properly taken from numerical studies in which the general-RMHD (GRMHD) effects are taken into account during the merger of the NSs.

\citet[][hereafter C17]{ciolfi2017} performed a 3D GRMHD numerical study of the coalescence of two NS for different EOS (soft: APR4, stiff: H4, and stiff: MS1) and mass ratios ($\rm{q}=0.9$ and 1.0). For the soft and equal mass case, the resultant density reaches values of $10^{14}-10^{8}$~g~cm$^{-3}$ and follows a $\rho \propto R^{-3}$ profile (with R the spherical radius), meanwhile the magnetic field (B) reaches $10^{16}-10^{12}$~G. \citet{Lazzati2021} and \citet{2021P} used the density morphology of different GRMHD simulations and they found that the jets propagating on a realistic medium present large deviations from the models that evolve in an isotropic and homogeneous medium. Non of the previous studies, though, studied the effects that a magnetized medium produces in the jet or its cocoon. Since the dynamical ejecta is vastly magnetized, it could modify the evolution of the relativistic jet which drills through it. The additional magnetic pressure could collimate the jet-cocoon even more and boost it or the magnetic material which is entrained into the cocoon and jet may have an important role. Additionally, the B field distribution could also affect the evolution of the jet. Thus it is important to study the effects that a magnetized medium produces in the evolution of a relativistic and collimated.

In this work, we present a comprehensive study of a series of 2.5 dimensional (2.5D) RMHD simulations in which a non-magnetized jet evolves through a magnetized medium (which resembles that from model APR4 with q1.0 of C17). The paper is organized as follows. In Section~\ref{s:setup} we  describe the setup and physics in our simulations. The results are presented in Section~\ref{s:results}. We discuss and conclude in Section~\ref{s:discussion}.

\section{{Global setup and models}}
\label{s:setup}
With the purpose of examining the evolution of a SGRB jet through a dense and magnetized medium produced by the the merger of two NSs, we followed a relativistic and collimated jet through a dense and magnetized medium similar to that from the GRMHD results of C17. Their simulations were carried out by using the relativistic ideal-magneto hydrodynamical code PLUTO \citep{mignone2012} in 2.5D spherical coordinates (i.e. a 2 dimensional system in which v$_{\rm{\phi}}$ and B$_{\rm{\phi}}$ were also followed).

Our initial conditions for the profiles of the density ($\rho_{\rm{m}}$) and thermal pressure ($P_{\rm{m}}$) were obtained from the C17 simulation of two equal mass NSs with soft EOS (model APR4). The medium was taken to be an ideal gas and the profiles were smoothed to follow $\rho_{\rm{m}} \propto R^{-3}$ (with $R$ the spherical radius), $P_{\rm{m}} \propto R^{-4}$ (with which the temperature follows a $R^{-1}$ profile). The total mass of the medium in the domain was $1.24\times10^{-2}$ M$_{\odot}$.

Since large-scale poloidal magnetic flux around the central engine can form due to the magnetorotational instability, and that persistent jets require poloidal magnetic flux on a large scale \citep{liska2020}, thus, we assumed that the B had a poloidal configuration and that its profile was $B \propto R^{-2}$ (thus $P_{\rm{B}} \propto R^{-4}$) in order to have a constant plasma beta ($\beta=P_{\rm_{g}}/P_{\rm{B}}$). The B magnitudes were consistent with those of C17 ($B_{\rm max} \sim 10^{16}-10^{15}$~G). In order to maintain $\nabla\cdot \bar{B}=0$ (and guarantee that the injected jet had velocities $v<c$), the 8-wave of \citet{Pow94} method was used. 

A non-magnetized, axisymmetric, relativistic, and collimated jet was followed as it drills through a dense and magnetized medium. The jet was injected as a boundary condition at a distance $d_{\rm co}=200$~km\footnote{This distance was chosen in order to avoid GR contributions.} from the center of the compact object. The jet was launched with a constant luminosity of $L_{\rm{j}}=2\times10^{50}$~erg~s$^{-1}$, an opening angle of $\theta_{j} = 10^{\circ}$, an initial Lorentz factor of $\Gamma_{\rm{j,0}}=5$, and a Lorentz factor at infinity of $\Gamma_{\rm{j,\infty}}=400$.

\begin{figure}
    \centering
    \includegraphics[width=\columnwidth]{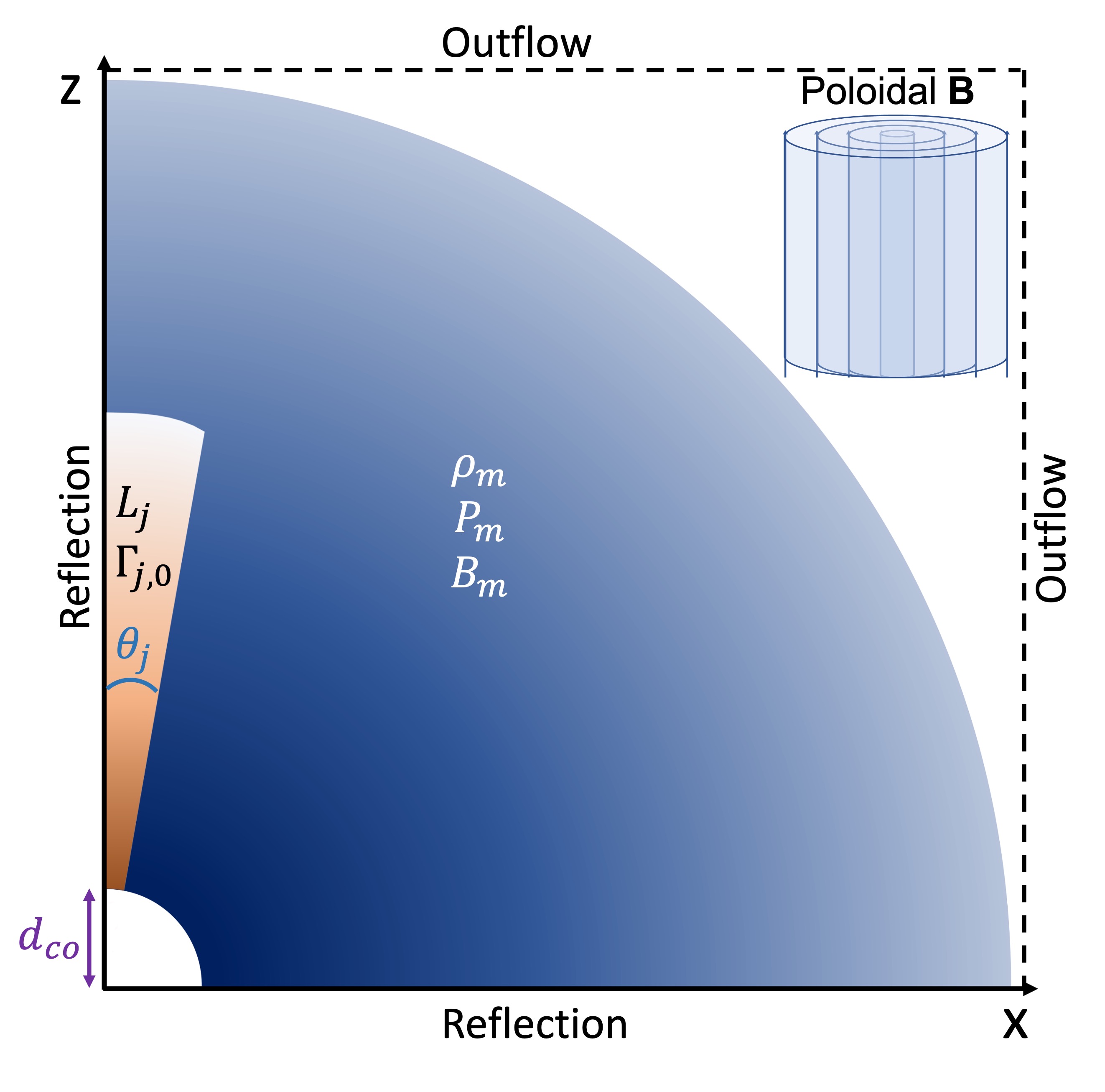}
    \caption{Scheme of the global setup. The figure shows the magnetized medium with $\rho_{\rm{m}}, P_{\rm{m}}$, and $B_{\rm{m}}$ through which a collimated and relativistic jet evolves. The jet (with $L_{\rm{j}}, \Gamma_{\rm{j,0}}$ and $\theta_{\rm{j}}$) is launched at a distance $d_{\rm co}$ from the compact object. The upper right figure shows the magnetic configuration (poloidal) of the medium. The boundary conditions are also shown.}
    \label{f:setup}
\end{figure}

Figure~\ref{f:setup} shows the global initial setup in spherical coordinates and the boundary conditions used in the simulations. 
The computational domain extended from r$_{\rm{min}}=2\times10^{7}$~cm to r$_{\rm{max}}=1.22\times10^{9}$~cm and between $\theta_{\rm{min}}=0$ and   $\theta_{\rm{max}}=\pi/2$. The jet was injected from r$_{\rm{min}}$ and $\theta_{\rm{min}}=0^{\circ}$ to $\theta=\theta_{\rm{j}}=10^{\circ}$; the remaining inner boundary had a reflective condition. The r$_{\rm{max}}$ boundary had a free outflow condition. The axisymmetric and equatorial boundaries had reflective conditions. All the simulations were carried out with a Courant number of $\rm{Co}=0.3$, and the total integration time was $t=0.11$~s. A spherical fixed mesh with N$_{r}$ radial divisions and N$_{\theta}$ angular divisions was used. The radial size of the cells increased logarithmically with an increment factor of 0.05$\%$, while the angular size of the cells was constant. Various resolutions were used in order to verify the convergence of the results. Specifically, we carried out four non-magnetized medium simulations at varying resolution to check for convergence (see below). In these, the low resolution (LR) had N$_{\rm{r}}=6000$ and N$_{\theta}=600$ cells, the medium resolution (MR) had N$_{\rm{r}}=8000$ and N$_{\theta}=800$ cells, the standard resolution (SR) used in the simulations had N$_{\rm{r}}=10000$ and N$_{\theta}=1000$ cells, and the high resolution (HR) had N$_{\rm{r}}=12000$ and N$_{\theta}=1200$ cells.

After we established the optimal resolution, we carried out calculations for  eleven models with distinct magnetization (parameterized through $\beta$), as well as a control model with no magnetization ($\beta = \infty$). The name of each model, the $\beta$ value, and the correspondent resolution are shown in Table~\ref{t:models}. In each model (PXX), P stands for poloidal, and the number corresponds to the value of $\beta$.

\begin{table}
\begin{center}
\begin{tabular}{ccc}
\hline
Name          & $\beta_{m}$    & Resolution       \\ 
\hline
Control       & $\infty$       & LR, MR, SR, HR         \\
P0.1          & 0.1            & SR                      \\
P0.5          & 0.5            & SR                      \\
P1.0          & 1.0            & SR                      \\
P5.0          & 5.0            & SR                      \\
P20           & 10             & SR                      \\
P25           & 25             & SR                      \\
P50           & 50             & SR                      \\
P75           & 75             & SR                      \\
P100          & 100            & SR                      \\
P500          & 500            & SR                      \\
P1e4          & $10^4$         & SR                      \\ 
\hline
\end{tabular}
\end{center}
\caption{Name, magnetization of the medium ($\beta_m$ value), and the resolution of each model. }
\label{t:models}
\end{table}

\section{Results}
\label{s:results}
\begin{figure*}
    \centering
   \includegraphics[width=0.95\textwidth]{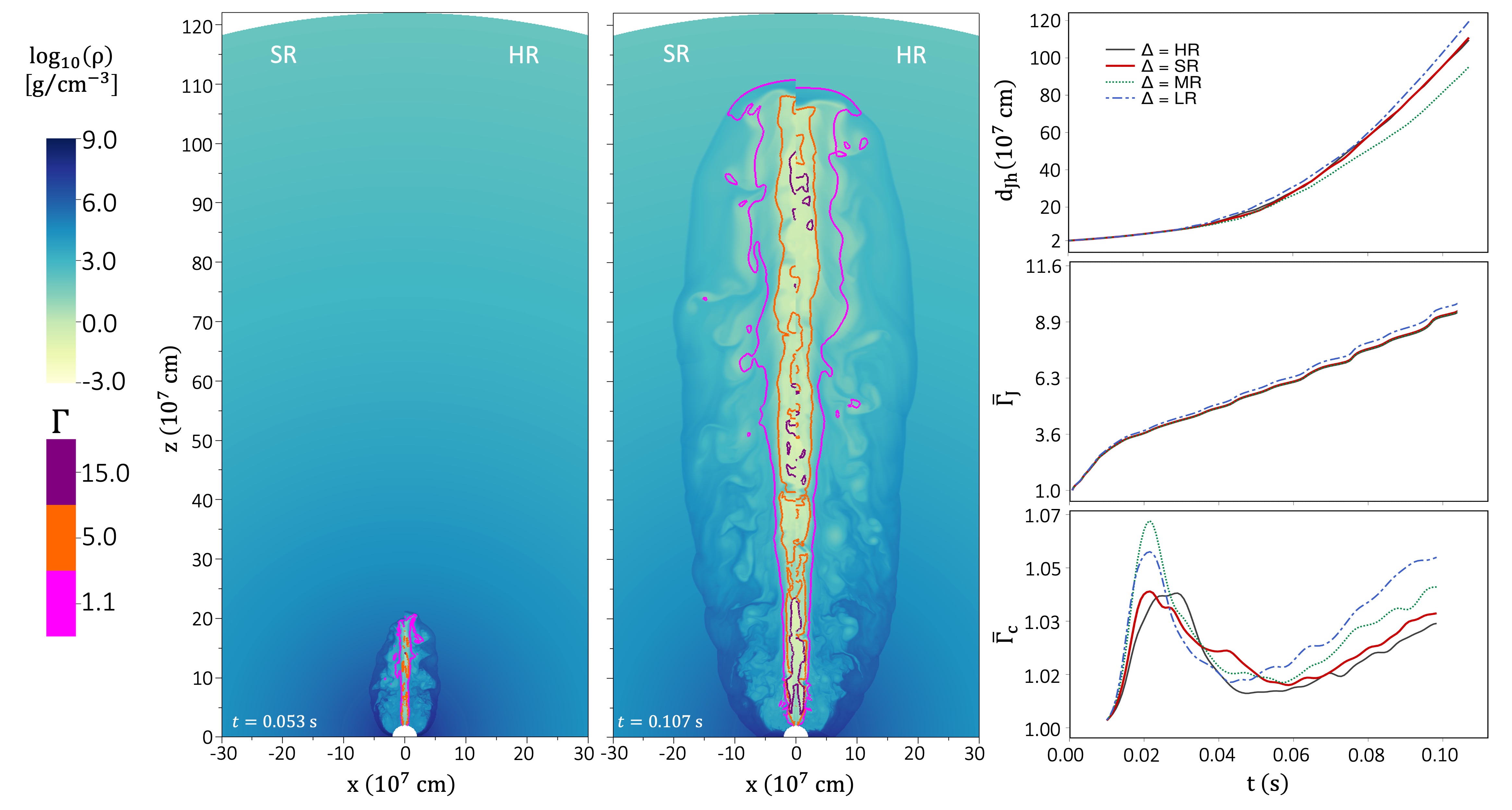}
    \caption{Density maps (in g cm$^{-3}$) of a relativistic jet drilling through a non-magnetized medium (control model) at two times ($t=0.053$~s, left panel; and $0.107$~s, right panel). The isocontours indicate the Lorentz factors ($\Gamma=1.1$ in pink, $\Gamma=5.0$ in orange, and $\Gamma=15.0$ in purple). In each time we show the result for the SR model (left) and HR model (right). The right-hand panels show the evolution of the jet-head ($d_{\rm{jh}}$), average Lorentz factor of the jet ($\overline{\Gamma}_{\rm j}$), and average Lorentz factor of the cocoon ($\overline{\Gamma}_{\rm c}$) for the four used resolutions (LR, MR, SR, and HR).}
    \label{f:Fig2P1}
\end{figure*}

In this section we describe the results of our calculations. We first ran a set of non-magnetized models with different resolutions (LR, MR, SR, and HR) to understand the effects of  resolution on the results. In Figure~\ref{f:Fig2P1} we present density maps (with different Lorentz factors shown as isocontours) in which the evolution of a relativistic jet through a non-magnetized medium is shown (at $t=0.053$~s, left panel; and 0.107~s, right panel). In both cases, we show the morphology of the jet and cocoon for both the SR (left-hand side) and the HR (right-hand side) cases. 
The comparison between the SR and HR shows that the general morphology of the jet is similar. At both epoch, the position of the head of the jet $d_{\rm{jh}}$ is almost coincident. Furthermore, the width of the cocoon does not change when the resolution is increased. The turbulence pattern inside of the cocoon is comparable for both models. Finally, the contours distribution of the Lorentz factor are analogous. At $t=0.053$~s, we found a 5.4$\%$  difference in $d_{\rm{jh}}$ (with the jet from the HR model being slightly larger). At $t=0.107$~s  $d_{\rm{jh}}$  for the SR model has reached a larger z-distance (1.2\% more than in the HR model). In both cases we found regions with $\Gamma>15$. The final morphology of the cocoon does not present significant differences between HR and SR models.

In the right panels of Figure~\ref{f:Fig2P1}, we present a convergence study that shows the reliability of our results. To evaluate the behavior of the different components we applied the method used in \citet{Lazzati2021} in which they divide the jet, cocoon, and ejecta material through the Lorentz factor at infinity\footnote{The asymptotic Lorentz factor that can be achieved if all internal energy is converted into bulk kinetic energy is found as: $\Gamma_{\rm{j}, \infty}=\Gamma_{\rm {j, 0}}\left( 1+4\frac{P_{\rm{j}}}{\rho_{\rm{j}} c^{2} } \right)$.}. We define as ejecta all the material with $1.0\leq\Gamma_{\infty} \leq 1.0001$ ($0<\beta_\infty<0.014$), as cocoon all the material with $1.0001\leq\Gamma_{\infty} \leq 5.0$, and as jet all the material with $\Gamma_{\infty} \geq 5.0$. The right-hand panels show the evolution of the jet-head ($d_{\rm jh}$, upper panel), the average Lorentz factor of the jet ($\overline{\Gamma}_{\rm{j}}$, middle panel), and the average Lorentz factor of the cocoon ($\overline{\Gamma}_{\rm c}$, lower panel) for the different resolution models. 
Comparing the LR, MR, and SR models we conclude that the evolution of the jet and cocoon varies as a function of the resolution (see for example how the average difference between the $d_{\rm{jh}}$, $\overline{\Gamma}_{\rm j}$, and $\overline{\Gamma}_{\rm c}$ between the LR and MR cases are 14.2\%, 4.2\%, 10.2\%, respectively). Therefore, convergence is not achieved in either the LR or MR models. Instead, comparing the SR and HR models we see that the evolution of the jet and cocoon are very similar. For the $d_{\rm{jh}}$ case the average difference between the HR and SR simulations is $1.5\%$. For the $\overline{\Gamma}_{\rm j}$ the average difference between the SR and HR is 3.7\%.
Finally, for the $\overline{\Gamma}_{\rm c}$, which presents variable behaviour (presumably due to the turbulence generated in the cocoon), the average difference between the SR and HR is less than 1\%. We use the SR as the resolution used in the rest of the magnetized models.

\begin{figure*}
    \centering
   \includegraphics[width=0.95\textwidth]{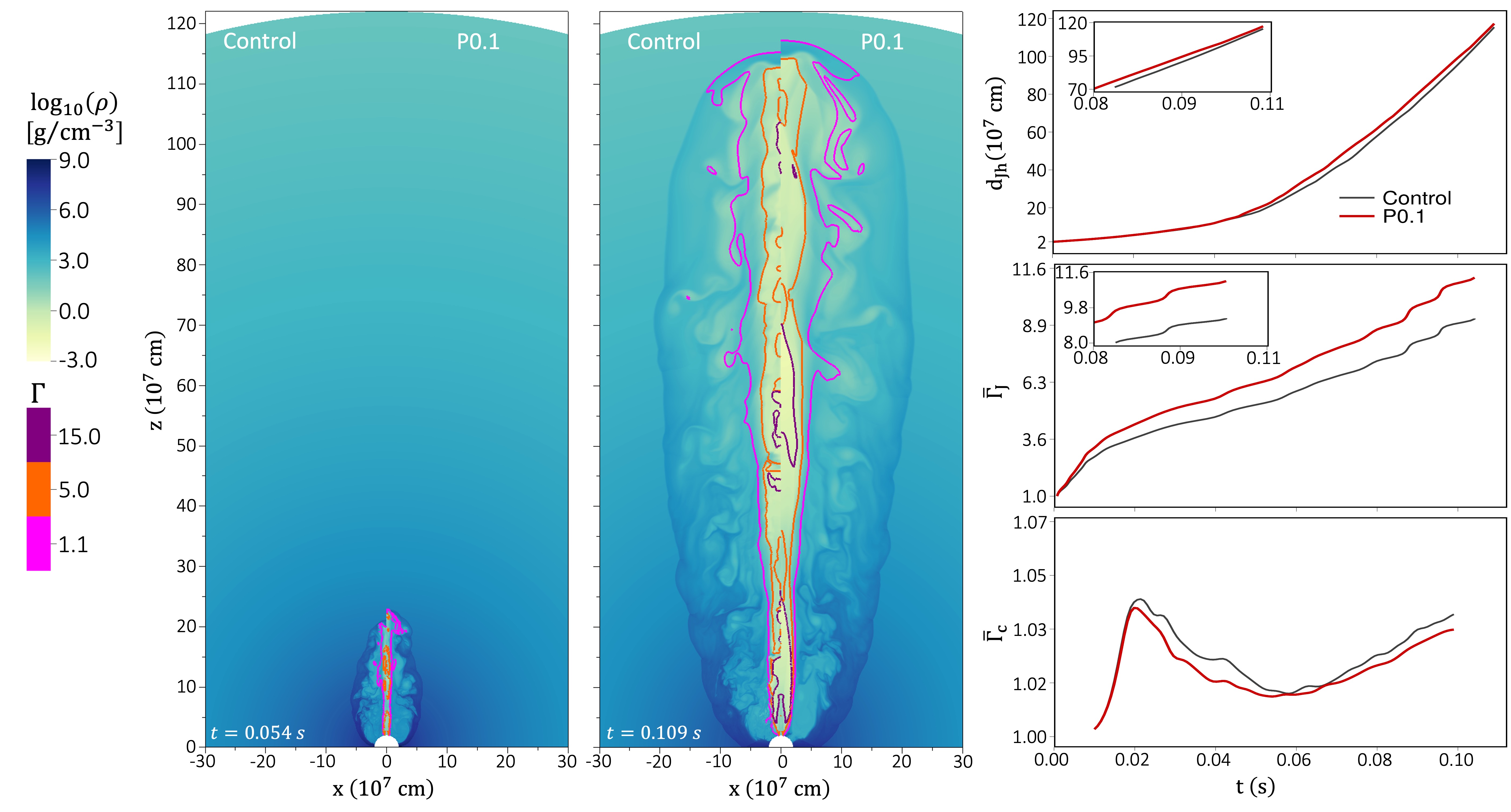}
    \caption{Same as Figure~\ref{f:Fig2P1}, for a jet drilling through a non-magnetized medium (control model) and a magnetized medium (model P0.1) at two times (0.054~s and 0.109~s). For the $d_{\rm{jh}}$ and $\overline{\Gamma}_{\rm j}$ evolution we also include insets of the zoom of the profiles.}
    \label{f:Fig3P1}
\end{figure*}

As an example of how the magnetic field affects the temporal evolution of a relativistic and collimated jet, we present in Figure~\ref{f:Fig3P1} a set of density maps showing the comparison of a relativistic jet drilling through a non-magnetized (left side of panels) and  a magnetized medium (right side of panels) at different times. Specifically, we show the jet at $t=0.054$~s (left-hand panel) and at 0.109~s (middle panel) drilling through a medium with $\beta=0.1$ (model P0.1). The Lorentz factor isocontours are the same as in Figure~\ref{f:Fig2P1}. The density maps show that at both times there are no marked differences in the evolution of the system. The cocoon height in the polar axis (i.e. at $\rm{x}=0$~cm) at $t=0.109$~s, for example, differs in both cases by less than 2\%. 
The right-hand panels show the evolution of  $d_{\rm{jh}}$, $\overline{\Gamma}_{\rm j}$, and $\overline{\Gamma}_{\rm c}$ in the non-magnetized (black line) and magnetized case (red line). The jet head distance and the average Lorentz factor of the cocoon are not significantly altered by the magnetization of the medium (see the profiles of $d_{\rm{jh}}$ and $\overline{\Gamma}_{\rm c}$ in Figure~\ref{f:Fig3P1}). The average difference for both $d_{\rm{jh}}$ and $\overline{\Gamma}_{\rm c}$ is less than 2\%. On the other hand, the average Lorentz factor of the jet is affected by the magnetic field as the jet moving through the magnetized medium is at all times higher than when it evolves in a non-magnetized medium.

\begin{figure*}
    \centering
    \includegraphics[height=0.55\textwidth ]{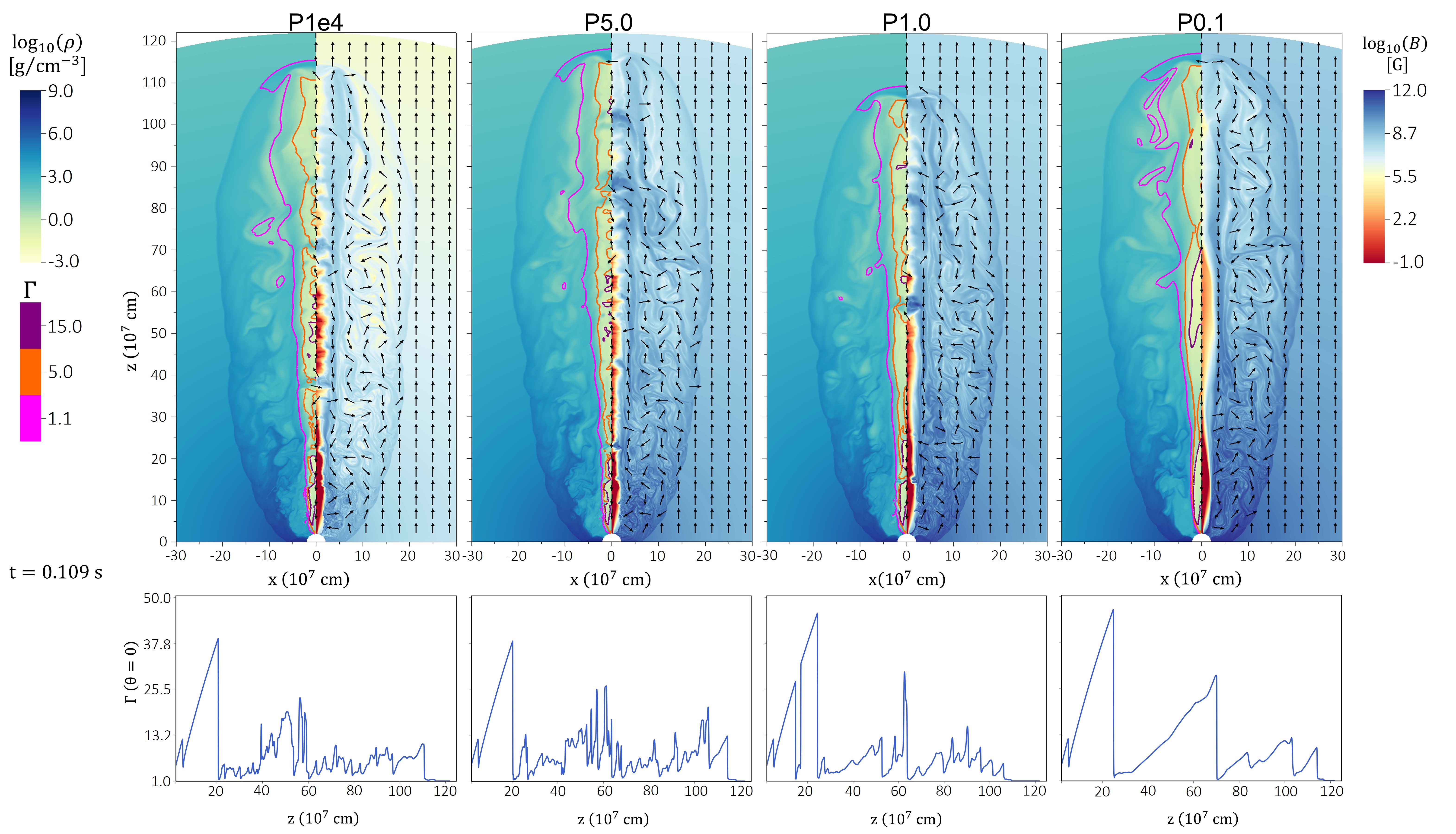}    
    \caption{Comparison of the evolution of a relativistic through differently magnetized media (all at $t=0.109$~s). From left to right we show models P1e4, P5.0, P1.0, and P0.1. The left-side of the upper panels present the density maps and Lorentz factor isocontours as in Figure~\ref{f:Fig2P1}. The right-side of the upper panels show the B(G) and the vector field flow. The lower panels display the Lorentz factor profile of the jet along the polar axis ($\Gamma_{\rm j}(r,\theta=0)$).}
    \label{f:Fig4P1}
\end{figure*}

To better understand the role of the ambient material magnetization, we carried out a set of simulations in which a relativistic and collimated jet evolved through media with different magnetization. Figure~\ref{f:Fig4P1} displays the configuration of the jet and its correspondent cocoon through media with $\beta=10^{4}, 5, 1, 0.1$ (models P1e4, P5.0, P1.0, and P0.1, respectively), all at $t=0.109$~s. The upper panels displays the density maps and Lorentz isocontours (left-hand side of each panel) as well as the magnitude of the magnetic field with the vector field lines (right-hand side of each panel). 
Globally, the cocoon is not drastically affected but the internal structure of the jet is. We do not find considerable differences in the morphology of the cocoon. The upper-left panels of each model shown how the Lorentz factor morphology is basically the same (independently of the magnetization). Also, the jet-head of all the magnetized models travel similar distances (though a $\sim$10\% variability may appear, for example, P1.0 and P5.0 which travelled the smallest and largest distances, respectively). Moreover, turbulent-like flows are present within the cocoon and an increase in the magnetization of the cocoon as a function of the medium's magnetic field is present. The cocoon of the lowest magnetized model P1e4 is basically the same as for the non-magnetized control case.
On the other hand, the jet (which in all cases is launched with no B field) is affected by the magnetized medium and gets magnetized (see how the front of the jet is more magnetized than its base). Model P0.1 presents the most magnetized jet at $t=0.109$~s. In order to appreciate how the magnetized medium affects the internal structure of the jet we show in the lower panels the Lorentz factor profile of the jet along the the polar axis ($\Gamma(\theta=0)$). Each panel corresponds to the model above it. The magnetization of the jet and cocoon has an impact on the jet structure since the low magnetized medium produces a jet with many RSs, and the highly magnetized medium produces a jet with fewer RSs.
Within the internal shock model (where the energy is dissipated between the RSs), the lack of shocks would produce a burst with a dim prompt emission and would eventually radiate the energy reservoir as a bright afterglow.
We finally note that the diffusion of the magnetic field in the cocoon is larger than in the jet, possibly due to the magnetic field that is trapped when the jet interacts with the medium.

\begin{figure*}
    \centering
    \includegraphics[width=0.95\textwidth]{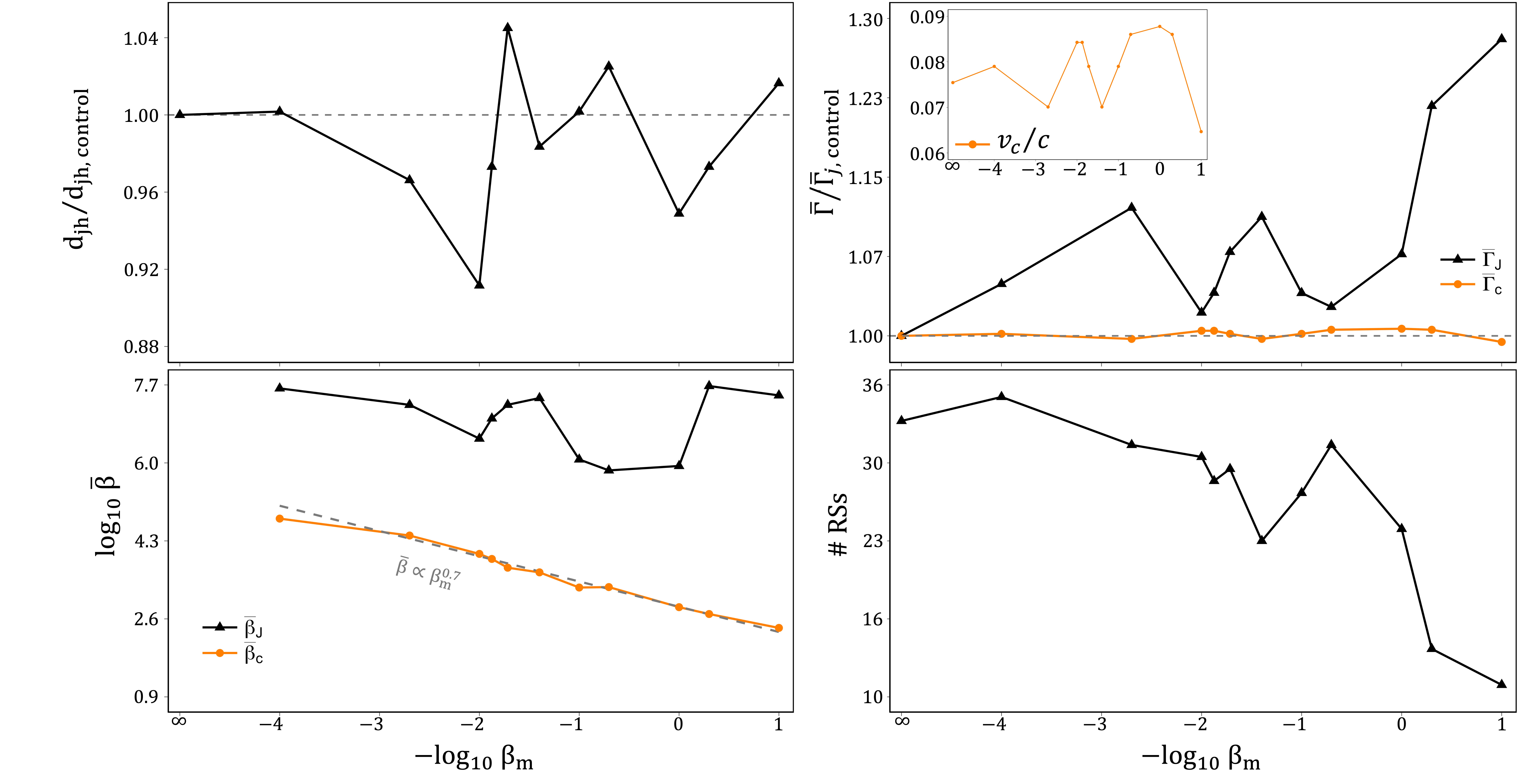}
    \caption{Quantitative analysis of the jet-head distance, average $\Gamma$ and $\beta$ of the jet and cocoon, and number of RSs for a jet moving through differently magnetized media ($\beta_m$). The upper-left panel displays the position of the jet head with respect to control model ($d_{\rm jh}/d_{\rm jh, control}$, the grey line indicates $d_{\rm jh}=d_{\rm jh, control}$). The top-right panel exhibit the relative average Lorentz factor of the jet $\overline{\Gamma}_{\rm j}/\overline{\Gamma}_{\rm j, control}$ (black line) and cocoon $\overline{\Gamma}_{\rm c}/\overline{\Gamma}_{\rm c, control}$ (orange line) whereas the inset panel presents the cocoon velocity in c units. The bottom-left panel shows the average beta value of the jet $\overline{\beta}_{\rm j}$ (black line) and cocoon $\overline{\beta}_{\rm c}$ (orange line). The bottom-right panel the number of RSs.}
    \label{f:Fig5P1}
\end{figure*}

To further understand the effects that the magnetization of the medium produces in the evolution of the jet-cocoon, we calculated $d_{\rm jh}$, average Lorentz factor of the jet and cocoon ($\overline{\Gamma}_{\rm j}$ and $\overline{\Gamma}_{\rm c}$, respectively), average $\beta$ for the jet and cocoon ($\overline{\beta}_{\rm j}$ and $\overline{\beta}_{\rm c}$, respectively), and the number of RSs in the jet for all the models with different magnetization $\beta_m$ (see Table~\ref{t:models} for details). The results are shown in Figure~\ref{f:Fig5P1}. 
The top-left panel shows that no trend is found for the $d_{\rm jh}$ as a function of $\beta_m$ and that it tends to be very close to the value of the non-magnetized model ($d_{\rm jh} \approx d_{\rm jh, control}$) whereas the inset panel exhibit the cocoon velocity in c units. However, there are small differences ($<10\%$) which are likely due to the energy dissipation produced by the turbulence in the cocoon.
The top-right panel shows that $\overline{\Gamma}_{\rm j}$ tends to increase with a higher magnetized medium (i.e. lower $\beta_m$), as already noticed in Figure~\ref{f:Fig3P1}. However, the inclusion of intermediate values of magnetization shows that there are local variations and that the trend is not monotonic. On the other hand, the average velocity of the material within the cocoon is basically the same for all models ($v_c/c \sim 0.06-0.09$, see inset of the correspondent panel), thus, $\overline{\Gamma}_{\rm c}$ is largely unaffected by the magnetization of the medium.

The bottom-left panel shows that the cocoon tends to be more magnetized when moving through media with higher B, this is, $\overline{\beta}_{\rm c}$ tends to decrease as a function of $\beta_m$. For $\overline{\beta}_{\rm j}$ we find no clear trend. From Figure~\ref{f:Fig4P1} we see that the base of the jet is mostly non-magnetized and the jet-head is magnetized, however the B magnitude of the jet is low, independently of $\beta_m$.
Finally, the bottom-right panel emphasizes the main difference among our simulations. There is a decreasing trend in the number of RSs in the jet as a function of magnetization (Figure~\ref{f:Fig4P1} also shows how the Lorentz factor at $\Gamma(\theta=0)$ is smoother in the most magnetized cases). The jet with a less magnetic field (model P1e4) presented 36 maxima compared with eleven peaks of model P0.1 (the most magnetized case).

\section{Discussion and conclusions}
\label{s:discussion}
In this work we studied the effects that a magnetized medium produces in the evolution of a relativistic jet that drills through it. A large set of models were followed using 2.5D RMHD simulations. The models set included a non-magnetized case and eleven cases with different magnetization. In addition, a quantitative study of the jet properties separating each component (cocoon and jet) was carried out. Our work shows that the ambient magnetic field, even at sub-dominant levels, can have a significant impact on the jet dynamics.

We find that in highly magnetized media smooth jets would be produced. Since the jets accelerate between shocks, a jet without many shocks has larger $\Gamma$ (see the upper right panel of Figure~\ref{f:Fig5P1}). The lack of shocks would produce a burst with a dim prompt emission and bright afterglow.

An interesting result is that even when the average Lorentz factor of the jet is boosted by the lack of RSs, the distance traveled by each jet is similar (see the upper panels of Figure~\ref{f:Fig5P1}). This can be because the jet head propagation depends mostly on the jet luminosity (see for example equations 4 and 5 in \citet{Matzner2003}). Therefore, the jet head velocity has very little dependence on the magnetic field intensity. In addition, we find that the magnetization of the cocoon follows that of the magnetized medium. Specifically, we find that $\beta_{c} \propto \beta_{c}^{0.7}$ (shown in the lower left panel of Figure~\ref{f:Fig5P1}), which is most likely because the cocoon is predominantly composed by shocked medium material. Meanwhile, the magnetization of the jet is largely unaffected by that of the medium. 

The light curve and polarization produced by the jet are key elements that may be compared with the observations. Since the domain which we used is small, thus, the jet-cocoon system is optically thick for all our models. It is necessary to evolve the jet up to at least the photospheric radius and use a radiative code, such as McRAT \citep{Lazzati2016}, to obtain these important features. Also, we must note that the presented work was limited to 2.5D, small domain, small integration times, non-magnetized and uniformly injected jets, and a poloidal B field in the medium. Three dimensional and bigger domains, variable jets that may be magnetized, and other B field distributions must be studied to fully understand the role of the magnetized medium in the evolution of the jet.


\section*{Acknowledgements}
We thank the referee for the helpful comments and improvement of the manuscript. 
LGG and DLC acknowledge the support from the Miztli-UNAM supercomputer (project LANCAD-UNAM-DGTIC-321) for the assigned computational time in which the simulations were performed and also thank the support of the UNAM-PAPIIT grant IG100820. 
DLC is supported by C\'atedras CONACyT at the Instituto de Astronom\'ia, UNAM. 
DL acknowledges support from NSF grant AST-1907955. LGG acknowledges support from CONACyT doctoral scholarship.
Many of the images in this study were produced using VisIt. VisIt is supported by the Department of Energy with funding from the Advanced Simulation and Computing Program, the Scientific Discovery through Advanced Computing Program, and the Exascale Computing Project.

\section*{Data Availability}
The data underlying this article will be shared on reasonable request to the corresponding author.



\appendix


\bsp	
\label{lastpage}

\begin{thebibliography}{99}

\bibitem[Abbott et al.(2017a)]{AbbotA2017} Abbott, B.~P., Abbott, R., Abbott, T.~D., et al.\ 2017 A, \apjl, 848, L12. doi:10.3847/2041-8213/aa91c9

\bibitem[Abbott et al.(2017b)]{AbbotB2017} Abbott, B.~P., Abbott, R., Abbott, T.~D., et al.\ 2017, \prl, 119, 161101. doi:10.1103/PhysRevLett.119.161101

\bibitem[Aloy et al.(2005)]{Aloy2005} Aloy, M.~A., Janka, H.-T., \& M{\"u}ller, E.\ 2005, \aap, 436, 273. doi:10.1051/0004-6361:20041865

\bibitem[\protect\citeauthoryear{Begelman \& Cioffi}{1989}]{Begelman1989} Begelman M.~C., Cioffi D.~F., 1989, ApJL, 345, L21. doi:10.1086/185542

\bibitem[Berger(2014)]{2014Ber} Berger, E.\ 2014, \araa, 52, 43. doi:10.1146/annurev-astro-081913-035926

\bibitem[\protect\citeauthoryear{Blandford \& Rees}{1974}]{Blandford1974} Blandford R.~D., Rees M.~J., 1974, MNRAS, 169, 395. doi:10.1093/mnras/169.3.395

\bibitem[\protect\citeauthoryear{Bo{\v{s}}njak, Daigne, \& Dubus}{2009}]{Bo2009} Bo{\v{s}}njak {\v{Z}}., Daigne F., Dubus G., 2009, A\&A, 498, 677. doi:10.1051/0004-6361/200811375

\bibitem[Bucciantini et al.(2008)]{Bucciantini2008} Bucciantini, N., Quataert, E., Arons, J., et al.\ 2008, \mnras, 383, L25

\bibitem[Bucciantini et al.(2009)]{Bucciantini2009} Bucciantini, N., Quataert, E., Metzger, B.~D., et al.\ 2009, \mnras, 396, 2038


\bibitem[Bromberg et al.(2018)]{Bromber2018} Bromberg, O., Tchekhovskoy, A., Gottlieb, O., et al.\ 2018, \mnras, 475, 2971. doi:10.1093/mnras/stx3316

\bibitem[\protect\citeauthoryear{Ciolfi et al.}{2017}]{ciolfi2017} Ciolfi R., Kastaun W., Giacomazzo B., Endrizzi A., Siegel D.~M., Perna R., 2017, PhRvD, 95, 063016. doi:10.1103/PhysRevD.95.063016

\bibitem[\protect\citeauthoryear{Costa et al.}{1997}]{costa97} Costa E., Frontera F., Heise J., Feroci M., in't Zand J., Fiore F., Cinti M.~N., et al., 1997, Natur, 387, 783. doi:10.1038/42885

\bibitem[\protect\citeauthoryear{Daigne \& Mochkovitch}{1998}]{Daigne1998} Daigne F., Mochkovitch R., 1998, MNRAS, 296, 275. doi:10.1046/j.1365-8711.1998.01305.x

\bibitem[Duffell et al.(2015)]{Duffell2015} Duffell, P.~C., Quataert, E., \& MacFadyen, A.~I.\ 2015, \apj, 813, 64. doi:10.1088/0004-637X/813/1/64

\bibitem[Duffell et al.(2018)]{Duffell2018} Duffell, P.~C., Quataert, E., Kasen, D., et al.\ 2018, \apj, 866, 3. doi:10.3847/1538-4357/aae084

\bibitem[Fong et al.(2015)]{2015Fong} Fong, W., Berger, E., Margutti, R., et al.\ 2015, \apj, 815, 102. doi:10.1088/0004-637X/815/2/102

\bibitem[Gehrels et al.(2009)]{2009Gh} Gehrels, N., Ramirez-Ruiz, E., \& Fox, D.~B.\ 2009, \araa, 47, 567. doi:10.1146/annurev.astro.46.060407.145147

\bibitem[Ghirlanda et al.(2018)]{Ghi2018} Ghirlanda, G., Nappo, F., Ghisellini, G., et al.\ 2018, \aap, 609, A112. doi:10.1051/0004-6361/201731598

\bibitem[\protect\citeauthoryear{Ghirlanda et al.}{2009}]{ghirlanda2009} Ghirlanda G., Nava L., Ghisellini G., Celotti A., Firmani C., 2009, A\&A, 496, 585. doi:10.1051/0004-6361/200811209

\bibitem[Ghirlanda et al.(2019)]{Ghirlanda2019} Ghirlanda, G., Salafia, O.~S., Paragi, Z., et al.\ 2019, Science, 363, 968. doi:10.1126/science.aau8815

\bibitem[\protect\citeauthoryear{Giannios \& Spruit}{2007}]{Giannios2007} Giannios D., Spruit H.~C., 2007, A\&A, 469, 1. doi:10.1051/0004-6361:20066739

\bibitem[Gill et al.(2019)]{Gill2019} Gill, R., Granot, J., De Colle, F., et al.\ 2019, \apj, 883, 15. doi:10.3847/1538-4357/ab3577

\bibitem[Goldstein et al.(2017)]{Goldstein2017} Goldstein, A., Veres, P., Burns, E., et al.\ 2017, \apjl, 848, L14. doi:10.3847/2041-8213/aa8f41

\bibitem[\protect\citeauthoryear{Gottlieb, Nakar, \& Piran}{2018}]{Got2018a} Gottlieb O., Nakar E., Piran T., 2018, MNRAS, 473, 576. doi:10.1093/mnras/stx2357

\bibitem[\protect\citeauthoryear{Gottlieb et al.}{2018}]{Got2018b} Gottlieb O., Nakar E., Piran T., Hotokezaka K., 2018, MNRAS, 479, 588. doi:10.1093/mnras/sty1462

\bibitem[Gottlieb et al.(2020)]{Got2020} Gottlieb, O., Bromberg, O., Singh, C.~B., et al.\ 2020, \mnras, 498, 3320. doi:10.1093/mnras/staa2567

\bibitem[Gottlieb et al.(2021)]{Got2021a} Gottlieb, O., Bromberg, O., Levinson, A., et al.\ 2021, \mnras. doi:10.1093/mnras/stab1068

\bibitem[Gottlieb \& Globus(2021)]{Got2021b} Gottlieb, O. \& Globus, N.\ 2021, arXiv:2105.01076

\bibitem[Granot et al.(2018)]{Granot2018} Granot, J., Gill, R., Guetta, D., et al.\ 2018, \mnras, 481, 1597. doi:10.1093/mnras/sty2308

\bibitem[Hallinan et al.(2017)]{Hallinan2017} Hallinan, G., Corsi, A., Mooley, K.~P., et al.\ 2017, Science, 358, 1579. doi:10.1126/science.aap9855

\bibitem[Hamidani et al.(2020)]{Hamidani2020} Hamidani, H., Kiuchi, K., \& Ioka, K.\ 2020, \mnras, 491, 3192. doi:10.1093/mnras/stz3231

\bibitem[Hamidani \& Ioka(2021)]{hamidani2021} Hamidani, H. \& Ioka, K.\ 2021, \mnras, 500, 627. doi:10.1093/mnras/staa3276

\bibitem[Harrison et al.(2018)]{Harrison2018} Harrison, R., Gottlieb, O., \& Nakar, E.\ 2018, \mnras, 477, 2128

\bibitem[\protect\citeauthoryear{Klebesadel, Strong, \& Olson}{1973}]{klebesadel73} Klebesadel R.~W., Strong I.~B., Olson R.~A., 1973, ApJL, 182, L85. doi:10.1086/181225

\bibitem[Kouveliotou et al.(1993)]{1993K} Kouveliotou, C., Meegan, C.~A., Fishman, G.~J., et al.\ 1993, \apjl, 413, L101. doi:10.1086/186969

\bibitem[Lamb \& Kobayashi(2018)]{Lamb2018} Lamb, G.~P. \& Kobayashi, S.\ 2018, \mnras, 478, 733. doi:10.1093/mnras/sty1108

\bibitem[\protect\citeauthoryear{Lazzati}{2016}]{Lazzati2016} Lazzati D., 2016, ApJ, 829, 76. doi:10.3847/0004-637X/829/2/76

\bibitem[\protect\citeauthoryear{Lazzati, Morsony, \& Begelman}{2009}]{Lazzati2009} Lazzati D., Morsony B.~J., Begelman M.~C., 2009, ApJL, 700, L47. doi:10.1088/0004-637X/700/1/L47

\bibitem[Lazzati et al.(2013)]{Davide2013} Lazzati, D., Morsony, B.~J., Margutti, R., et al.\ 2013, \apj, 765, 103

\bibitem[Lazzati et al.(2017)]{Lazzati2017} Lazzati, D., Deich, A., Morsony, B.~J., et al.\ 2017, \mnras, 471, 1652. doi:10.1093/mnras/stx1683

\bibitem[Lazzati et al.(2018)]{Lazzati2018} Lazzati, D., Perna, R., Morsony, B.~J., et al.\ 2018, \prl, 120, 241103. doi:10.1103/PhysRevLett.120.241103

\bibitem[Lazzati \& Perna(2019)]{LazzatiPerna} Lazzati, D. \& Perna, R.\ 2019, \apj, 881, 89. doi:10.3847/1538-4357/ab2e06

\bibitem[Lazzati et al.(2020)]{Lazzati2020} Lazzati, D., Ciolfi, R., \& Perna, R.\ 2020, \apj, 898, 59. doi:10.3847/1538-4357/ab9a44

\bibitem[Lazzati et al.(2021)]{Lazzati2021} Lazzati, D., Perna, R., Ciolfi, R., et al.\ 2021, \apjl, 918, L6. doi:10.3847/2041-8213/ac1794

\bibitem[\protect\citeauthoryear{Liska, Tchekhovskoy, \& Quataert}{2020}]{liska2020} Liska M., Tchekhovskoy A., Quataert E., 2020, MNRAS, 494, 3656. doi:10.1093/mnras/staa955

\bibitem[Margutti et al.(2017)]{Margutti2017} Margutti, R., Berger, E., Fong, W., et al.\ 2017, \apjl, 848, L20. doi:10.3847/2041-8213/aa9057

\bibitem[Matzner(2003)]{Matzner2003} Matzner, C.~D.\ 2003, \mnras, 345, 575. doi:10.1046/j.1365-8711.2003.06969.x

\bibitem[\protect\citeauthoryear{M{\'e}sz{\'a}ros}{2019}]{meszaros2019} M{\'e}sz{\'a}ros P., 2019, MmSAI, 90, 57

\bibitem[\protect\citeauthoryear{Mignone et al.}{2012}]{mignone2012} Mignone A., Zanni C., Tzeferacos P., van Straalen B., Colella P., Bodo G., 2012, ApJS, 198, 7. doi:10.1088/0067-0049/198/1/7

\bibitem[Mooley et al.(2018)]{Mooley2018} Mooley, K.~P., Deller, A.~T., Gottlieb, O., et al.\ 2018, \nat, 561, 355. doi:10.1038/s41586-018-0486-3

\bibitem[Murguia-Berthier et al.(2014)]{Murguia2014} Murguia-Berthier, A., Montes, G., Ramirez-Ruiz, E., et al.\ 2014, \apjl, 788, L8. doi:10.1088/2041-8205/788/1/L8

\bibitem[Murguia-Berthier et al.(2017)]{Murguia2017} Murguia-Berthier, A., Ramirez-Ruiz, E., Montes, G., et al.\ 2017, \apjl, 835, L34. doi:10.3847/2041-8213/aa5b9e

\bibitem[Murguia-Berthier et al.(2021)]{Murguia2021} Murguia-Berthier, A., Ramirez-Ruiz, E., De Colle, F., et al.\ 2021, \apj, 908, 152. doi:10.3847/1538-4357/abd08e

\bibitem[Nagakura et al.(2014)]{Nagakura2014} Nagakura, H., Hotokezaka, K., Sekiguchi, Y., et al.\ 2014, \apjl, 784, L28

\bibitem[Nakar et al.(2018)]{Nakar2018} Nakar, E., Gottlieb, O., Piran, T., et al.\ 2018, \apj, 867, 18. doi:10.3847/1538-4357/aae205

\bibitem[Pavan et al.(2021)]{2021P} Pavan, A., Ciolfi, R., Kalinani, J.~V., et al.\ 2021, \mnras, 506, 3483. doi:10.1093/mnras/stab1810

\bibitem[\protect\citeauthoryear{Pe'er, M{\'e}sz{\'a}ros, \& Rees}{2006}]{Peer2006} Pe'er A., M{\'e}sz{\'a}ros P., Rees M.~J., 2006, ApJ, 642, 995. doi:10.1086/501424

\bibitem[Piran(1999)]{fireball} Piran, T.\ 1999, \physrep, 314, 575. doi:10.1016/S0370-1573(98)00127-6

\bibitem[Powell(1994)]{Pow94} Powell, K.~G.\ 1994, 

\bibitem[\protect\citeauthoryear{Rees \& M{\'e}sz{\'a}ros}{2005}]{Rees2005} Rees M.~J., M{\'e}sz{\'a}ros P., 2005, ApJ, 628, 847. doi:10.1086/430818

\bibitem[\protect\citeauthoryear{Rouco Escorial et al.}{2022}]{rouco2022} Rouco Escorial A., Fong W.-. fai ., Berger E., Laskar T., Margutti R., Schroeder G., Rastinejad J.~C., et al., 2022, arXiv, arXiv:2210.05695

\bibitem[\protect\citeauthoryear{Ryde et al.}{2011}]{Ryde2011} Ryde F., Pe'er A., Nymark T., Axelsson M., Moretti E., Lundman C., Battelino M., et al., 2011, MNRAS, 415, 3693. doi:10.1111/j.1365-2966.2011.18985.x

\bibitem[\protect\citeauthoryear{Scheuer}{1974}]{Scheuer1974} Scheuer P.~A.~G., 1974, MNRAS, 166, 513. doi:10.1093/mnras/166.3.513

\bibitem[Sharan Salafia \& Ghirlanda(2022)]{om2022} Sharan Salafia, O. \& Ghirlanda, G.\ 2022, arXiv:2206.11088

\bibitem[Urrutia et al.(2021)]{Ger2021} Urrutia, G., De Colle, F., Murguia-Berthier, A., et al.\ 2021, \mnras, 503, 4363. doi:10.1093/mnras/stab723


\bibitem[Troja et al.(2017)]{Troja2017} Troja, E., Piro, L., van Eerten, H., et al.\ 2017, \nat, 551, 71. doi:10.1038/nature24290

\bibitem[Xie et al.(2018)]{Xie2018} Xie, X., Zrake, J., \& MacFadyen, A.\ 2018, \apj, 863, 58. doi:10.3847/1538-4357/aacf9c

\end{thebibliography}
\end{document}